\pdfoutput=1 

\documentclass[conference]{IEEEtran}
\usepackage{cite}
\usepackage{threeparttable}
\usepackage{amsmath,amssymb,amsfonts}
\usepackage{enumitem}
\ifCLASSOPTIONcompsoc
\usepackage[caption=false,font=normalsize,labelfon
t=sf,textfont=sf]{subfig}
\else
\usepackage[caption=false,font=footnotesize]{subfig}
\fi
\usepackage{multirow}
\usepackage[super]{nth}
\usepackage{layouts}

\usepackage{todonotes}

\usepackage{siunitx}
\usepackage{acronym}
\acrodef{pe}[PE]{printed electronics}

\newcommand{\vthetha}{\ensuremath{\mathrm{\boldsymbol{\theta}}}}

\usepackage{algorithmic}
\usepackage{graphicx}
\usepackage{textcomp}
\usepackage{xcolor}
\usepackage{pifont}
\usepackage{makecell}
\def\BibTeX{{\rm B\kern-.05em{\sc i\kern-.025em b}\kern-.08em
    T\kern-.1667em\lower.7ex\hbox{E}\kern-.125emX}}

\newcommand{\red}[1]{{\color{black}#1}}
\newcommand{\blue}[1]{{\color{black}#1}}

\usepackage{flushend}
\newcommand{\textapprox}{\raisebox{0.5ex}{\texttildelow}}

\begin{document}
\bstctlcite{IEEEexample:BSTcontrol} 
\setlength{\abovedisplayskip}{2ex}
\setlength{\belowdisplayskip}{2ex}


\title{{\vspace{-1.5cm}\small To appear at the \textit{27th Design, Automation, and Test in Europe Conference (DATE'24)}, Valencia, Spain, 2024. \\
DOI: 10.23919/DATE58400.2024.10546879\\ 
}\vspace{-0.8\baselineskip}
\rule{\textwidth}{0.4pt}
Embedding Hardware Approximations in Discrete Genetic-based Training for Printed MLPs
}

\author{\IEEEauthorblockN{
Florentia Afentaki\IEEEauthorrefmark{1}\IEEEauthorrefmark{3},
Michael Hefenbrock\IEEEauthorrefmark{2},
Georgios Zervakis\IEEEauthorrefmark{1},
Mehdi B. Tahoori\IEEEauthorrefmark{3}
}
\IEEEauthorblockA{
\IEEEauthorrefmark{1}University of Patras, Greece,
\IEEEauthorrefmark{2}RevoAI GmbH, Germany,
\IEEEauthorrefmark{3}Karlsruhe Institute of Technology, Germany,
}
\IEEEauthorblockA{
\IEEEauthorrefmark{1}\{afentaki, zervakis\}@ceid.upatras.gr,
\IEEEauthorrefmark{2}\{michael.hefenbrock\}@revoai.de,
\IEEEauthorrefmark{3}\{mehdi.tahoori\}@kit.edu
}
}

\maketitle
\begin{abstract}
Printed Electronics (PE) stands out as a promising technology for widespread computing due to its distinct attributes, such as low costs and flexible manufacturing. 
Unlike traditional silicon-based technologies, PE enables stretchable, conformal, and non-toxic hardware. 
However, PE are constrained by larger feature sizes, making it challenging to implement complex circuits such as machine learning (ML) classifiers.
Approximate computing has been proven to reduce the hardware cost of ML circuits such as Multilayer Perceptrons (MLPs).
In this paper, we maximize the benefits of approximate computing by integrating hardware approximation into the MLP training process.
Due to the discrete nature of hardware approximation, we propose and implement a genetic-based, approximate, hardware-aware training approach specifically designed for printed MLPs.
For a 5\% accuracy loss, our MLPs achieve over 5$\times$ area and power reduction compared to the baseline while outperforming state-of-the-art approximate and stochastic printed MLPs.

\end{abstract}



\section{Introduction}\label{sec:intro}


Printed electronics (PE) offer a promising direction for integrating computing and intelligence across domains like smart bandages, disposable items, packaged goods, and smart packaging. 
These applications, including in-situ monitoring and the Fast-Moving Consumer Goods market, have specific fabrication cost, conformity, and time-to-market needs that traditional silicon-based electronics struggle to meet~\cite{Bleier:ISCA:2020:printedmicro}.

PE~\cite{Mubarik:MICRO:2020:printedml} enables these applications through additive manufacturing processes, producing conformal, low-cost hardware on demand. 
However, PE cannot match silicon VLSI systems in density, area, or speed due to larger feature sizes from imprecise printing. 
PE circuits operate at frequencies ranging from a few $\si{\hertz}$ to a few $\si{\kilo\hertz}$~\cite{cadilha2017digital}, with micrometer-sized features~\cite{lei2019low}. 
The large feature size and the capacitance in PE technology raise area and power consumption, deterring conventional digital architectures, like Machine~Learning~(ML) classifiers~\cite{Mubarik:MICRO:2020:printedml} which are the primary printed applications~\cite{Armeniakos:TC2023:codesign}.

As an attempt to address the above limitations, the authors in~\cite{Mubarik:MICRO:2020:printedml} leverage the customization potential offered by low-fabrication and Non-Recurring Engineering costs associated with printed circuits through bespoke circuit designs.
The term ``bespoke'' refers to fully-customized circuits tailored to specific ML model and dataset.
While~\cite{Mubarik:MICRO:2020:printedml} successfully reduced area and power for simple ML algorithms, the hardware overheads remain prohibitive for more complex algorithms like Multilayer~Perceptron~(MLP).
Table~\ref{tab:baselines} presents the hardware cost of several printed MLP circuits that follow the bespoke design of~\cite{Mubarik:MICRO:2020:printedml}.
As shown, all MLPs feature excessive power consumption and cannot be powered by any available printed battery~\cite{Mubarik:MICRO:2020:printedml, Armeniakos:DATE2022:axml} while their area requirement is above $12\si{\square\centi\meter}$, thus being unsuitable for the most printed applications~\cite{Armeniakos:TC2023:codesign}.

Targeting printed MLPs,~\cite{Armeniakos:DATE2022:axml,Armeniakos:TCAD2023:cross,Armeniakos:TC2023:codesign,Kokkinis:DATE2023} employed Approximate Computing (AxC) exploiting the high error resilience of ML applications~\cite{Henkel:ICCAD2022:expedition}.
Leveraging that in bespoke ML circuits the model's coefficients are hardwired and thus determine the circuit's area,~\cite{Armeniakos:DATE2022:axml} and~\cite{Armeniakos:TC2023:codesign} replace the MLP's coefficients with more area-efficient values reducing the multipliers' area.
Additionally, to reduce the cost of additions,~\cite{Armeniakos:DATE2022:axml} applied gate-level pruning while~\cite{Armeniakos:TC2023:codesign} used truncation in accumulations. 
Armeniakos et al.~\cite{Armeniakos:TCAD2023:cross}, extend~\cite{Armeniakos:DATE2022:axml} by applying voltage over-scaling.
In all the above works, multipliers are still required and thus the obtained gains are limited. 
In contrast,~\cite{Weller:2021:printed_stoch} designed stochastic printed MLPs but resulted in poor accuracy.

Prior works focus on employing AxC post-training, often compromising the balance between area and accuracy, thereby yielding sub-optimal results.
In this work, we address this limitation by incorporating hardware approximation during MLP training.
Gradient-based learning (backpropagation) relies on the loss function differentiability. 
Unfortunately, hardware approximations often involve discrete variables and decisions, which do not permit computing the gradients~\cite{Armeniakos:AxDNNsurvey}.
Additionally, to achieve hardware-aware training, both accuracy and area overhead must be addressed as objectives, framing the training process as a multi-objective optimization problem. 
Traditional gradient-based learning methods are not directly applicable to multi-objective optimization and often necessitate to be transformed to a single-objective optimization problem~\cite{benmeziane:arxiv2021:hw_awareNAS}. 
Therefore, we employ evolutionary methods like a Genetic Algorithm (GA), which, due to their capability to work with discrete variables, enable us to fully leverage hardware AxC techniques during our training.
AxC circuits' inherent attributes and evolutionary circuit design principles are expected to synergize positively~\cite{sekanina:TEvolutioryC:2015:evolution_to_ax_circuits}. 

We propose a GA-based approach for training hardware and approximation-aware bespoke printed MLPs.
While evolutionary approaches can be time-consuming for training large neural networks, for the small MLPs typically used in printed electronics~\cite{Mubarik:MICRO:2020:printedml,Armeniakos:TC2023:codesign}, GA-based training offers a promising alternative.
It allows highly optimized bespoke MLP circuits by incorporating discrete hardware-aware approximations during training, enabling simultaneous optimization of accuracy and hardware costs within a reasonable time frame.
For approximations, we adopt the state-of-the-art power-of-two~(pow2) weight quantization, to eliminate multiplications, and a finer-grained unstructured pruning, implemented through Full-Adders~(FAs) removal, to a approximate additions.

To the best of our knowledge, this is the first time that such a framework\footnote{{\scriptsize https://github.com/floAfentaki/Approximation-Techniques-Targeting-Printed-MLPs}} is proposed for hardware-efficient printed MLP circuits.
Our experiments across various MLPs show that our framework reduces both area and power by over \red{$5\times$} compared to the exact baseline and outperforms the current state-of-the-art approximate works.
Notably, it enables printed-energy-harvester operation for the majority of the examined MLPs.

\begin{table}[t!]
\setlength\tabcolsep{3pt}
\caption{Evaluation of the baseline printed MLPs
}
\label{tab:baselines}
\footnotesize
\centering
\renewcommand{\arraystretch}{1}
\begin{threeparttable}
\begin{tabular}{|l|cc|ccc|}
\cline{2-6}
\multicolumn{1}{c|}{} &  \multicolumn{5}{c|}{\textbf{Baseline}} \\ \hline
\multicolumn{1}{|c|}{\textbf{MLP}} & \textbf{Topology\tnote{1}} & \textbf{Parameters}\tnote{2}  & \textbf{Acc}\tnote{3}    & \begin{tabular}[c]{@{}c@{}}\textbf{Area} \\ ($\si{\square\centi\meter}$)\end{tabular} & \begin{tabular}[c]{@{}c@{}}\textbf{Power}\\ ($\si{\milli\watt}$)\end{tabular}  \\ \hline
\textbf{Breast Cancer}  & (10,3,2)  &38   & 0.980  & 12.0 & 40.0 \\
\textbf{Cardio}         & (21,3,3)  &78   & 0.881  & 33.4 & 124 \\
\textbf{Pendigits}      & (16,5,10) &145  & 0.937  & 67.0 & 213 \\
\textbf{RedWine}        & (11,2,6)  &42   & 0.564  & 17.6 & 73.5\\
\textbf{WhiteWine}      & (11,4,7)  &83   & 0.537  & 31.2 & 126 \\ \hline
\end{tabular}
\begin{tablenotes}\footnotesize
\vspace{0.5ex}
\item[] 
$^1$MLP topology. $^2$MLP parameters. $^3$Accuracy.
\vspace{-3.5ex}
\end{tablenotes}
\end{threeparttable}
\end{table}



\section{Background on Printed Electronics}\label{sec:background}

PE employs various printing methods such as jet, screen, or gravure printing~\cite{cui2016printed}.
These printing techniques are characterized by being mask-less, portable, and additive in nature, leading to significant reductions in manufacturing costs and production timelines~\cite{chang2017circuits}.
PE technology is being categorized into two main manufacturing approaches: additive and subtractive processes.
The additive manufacturing process consists of deposition steps, where the functional materials are directly deposited on the substrate, while the subtractive process integrates both additive and subtractive stages, similar to methodologies seen in silicon-based techniques~\cite{Henkel:ICCAD2022:expedition}.

The simplicity and low equipment costs of additive manufacturing process, enable the production of remarkably low-cost electronic circuits, even sub-cent levels. 
However, this process is characterized by low precision fabrication which results in increased device latency and
low integration density compared to silicon VLSI systems~\cite{Henkel:ICCAD2022:expedition}.
Though, target applications pose relaxed frequency and computational precision requirements, making printed circuits viable~\cite{Henkel:ICCAD2022:expedition}.
We focus on the Electrolyte-Gated FET (EGFET) technology, which features low supply voltage ($\leq1\si{\volt}$) and good mobility characteristics, making it well-suited for battery-powered applications~\cite{Bleier:ISCA:2020:printedmicro}.


\section{Approximate Computing Techniques}\label{subsec:ax_methods}

AxC promises significant area and power gains at the cost of some accuracy loss.
AxC has found widespread use in ML applications, particularly due to the increased computational demands and the inherent approximate nature in ML models~\cite{Henkel:ICCAD2022:expedition}.
Considering that in printed ML circuits feasibility is the fundamental concern, preceding the need for high accuracy, in our work we design approximate printed MLPs targeting to minimize the associated hardware overheads.
Hereafter, we briefly discuss the selection and hardware impact of the approximation techniques adopted in our work.

\subsection{Mutliplier Approximation}\label{subsubsec:quantization}

The primary contributor to area consumption within a neuron is the multiplier, closely followed by the accumulator~\cite{Armeniakos:AxDNNsurvey}. 
Consequently, the current state-of-the-art approximate printed ML approaches~\cite{Armeniakos:DATE2022:axml,Armeniakos:TC2023:codesign,Armeniakos:TCAD2023:cross} primarily target to diminish the multipliers area.
Nevertheless, despite the efforts made, multipliers are still required in~\cite{Armeniakos:DATE2022:axml,Armeniakos:TC2023:codesign,Armeniakos:TCAD2023:cross}, consuming considerable area.
To address this, we propose to take a step further and implement multiplier-less neurons by adopting the well-known approximation of quantizing the weights as powers-of-two.
As a result, each weight $w^{(l)}_{i,j}$, of the $i$-th input inside the $j$-th neuron of the $l$-th layer, is written as:
\begin{equation}
    w^{(l)}_{i,j} = s^{(l)}_{i,j}\cdot 2^{k^{(l)}_{i,j}} \quad  \text{with} \quad k^{(l)}_{i,j} \in [0,n-1), 
    \label{eq:weight}
\end{equation}
where $s^{(l)}_{i,j} \in \{-1, +1\}$ is the sign of the weight, and $n$ determines the number of bits used for weight representation.
Assuming a bespoke implementation, as the design standard in PE~\cite{Mubarik:MICRO:2020:printedml,Armeniakos:TC2023:codesign}, multiplication by a constant pow2 weight requires only wiring.
Hence, the cost of multipliers is effectively nullified.
The sign $s$ determines whether the product (input by pow2 weight) will be added or subtracted.
Still, since the inputs are always positive (QReLU activation is used) and the weight sign is also fixed, the product will always be added or always subtracted.
Therefore, if $s=-1$ only wiring and a few \texttt{NOT} gates are required since the `$1$' from all two's complement negations may be accumulated in the constant bias term before design, leading to no additional overhead.
As typical, we also use low-bitwidth quantized biases denoted $b^{(l)}_{j}$.


\subsection{Adder approximation}\label{subsec:axadd}
After eliminating the multipliers, the neuron's area is mainly determined by the area of the multi-operand adder.
Unstructured pruning is a widely used technique to compress a model and reduce its complexity.
Unstructured pruning removes connections and thus, in our printed MLPs where multipliers are already removed, unstructured pruning would directly remove a summand from the addition circuit. 
However, in small MLPs, like the targeted ones, where only a few weights are considered, pruning can lead to unacceptable accuracy loss while yielding only moderate hardware gains~\cite{Kokkinis:DATE2023}.

To address this, we adopt a more fine-grained approach to unstructured pruning, to strike a better balance between accuracy and the reduction of adder area.
Instead of completely removing an entire connection, we selectively eliminate only certain bits.
For example, if an input activation is a $6$-bit binary signal \mbox{${A=a_5a_4a_3a_2a_1a_0}_{_2}$}, instead of nullifying the entire signal~$A$, we nullify some of its bits, e.g., 
${A^\prime=a_50a_3a_20a_0}_{_2}$.
Again, since we design bespoke MLPs, accumulating multiple predefined `$0$' values in the same column of the adder tree will decrease the number of full-adders~(FAs) required and potentially reduce its height.
For instance, for every three constant `$0$' in a column, one FA is eliminated from that column and one carry connection to the right column is also removed.
Similarly to unstructured pruning our approach can also be implemented using masking.
For each weight $w^{(l)}_{i,j}$ we need to identify a mask $m^{(l)}_{i,j}$.
For each `$1$' bit in $m^{(l)}_{i,j}$, the respective bit of the input activation is retained for the addition, while for each `$0$' bit in $m^{(l)}_{i,j}$, the respective bit of the input activation is removed.
Assuming that \red{$x^{(l)}_i$} is the input activation of the $l$-th layer, which is multiplied with $w^{(l)}_{i,j}$, then the mask is applied to the input $x^{(l)}_i$ as $x^{(l)}_i \odot m^{(l)}_{i,j}$, where \(\odot\) is the the bitwise \texttt{AND} operation.
Thus, the mask of the previous example is \mbox{$m=101101_2$} and $A^\prime=A\odot m$.
If a mask is zero, the entire summand is removed.
Therefore, in~\eqref{eq:weight}, there's no need to define a zero weight, as it is hardware-equivalent to a zero mask.
The size of the masks depends on the size of the input features of each layer.
Moreover, we use the QReLU activation.
Unlike ReLU activation, which yields unbounded outputs, QReLU effectively limits the size of its output, resulting in smaller required bitwidths.
For our analysis, we use $4$ bits for the inputs and $8$ bits for the QReLU output.
These values are small enough and result in almost no accuracy degradation compared to larger bitwidths.
Hence, only small integer values are required to represent each mask $m^{(l)}_{i,j}$.
Notably, the masks are used solely for the high-level representation of the applied approximation, e.g., training.
In terms of hardware, there is no need for an \texttt{AND} gate for masking; the masked bits are directly removed from the adder.


\begin{figure}[!t]
\centering
\includegraphics[width=\columnwidth]{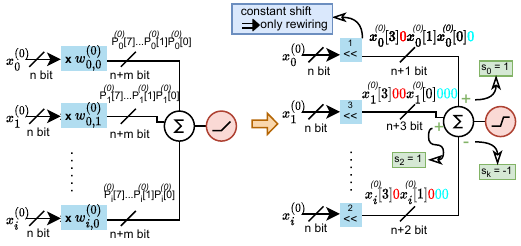}
\vspace{-3.5ex}
\caption{Showcase of the approximate neuron.
The figure on the left and right present the bespoke neuron before and after the hardware approximations.}
\label{fig:ax:neuron}
\vspace{-3ex}
\end{figure}

\subsection{Approximate Neuron}\label{subsec:axneuron}

An illustrative example of applying the aforementioned approximation techniques are depicted in \figurename~\ref{fig:ax:neuron}.
For readability, this example assumes 4-bit activations. 
As shown in \figurename~\ref{fig:ax:neuron}, multiplication is achieved through direct wiring of inputs, while addition approximation involves hardcoding zeros in the summand description, and the sign is also hardcoded.

As aforementioned, the primary factor influencing an approximate MLP's area is the accumulations involving multi-operand adder trees. 
Thus, a straightforward estimate for the MLP's area is the summation of the areas of these adder trees:
\begin{equation}\label{eq:area}
\operatorname{Area}(\vthetha) =
\sum_{\forall l,j}{\operatorname{AdderArea}(\mathbf{\vthetha}^{(l)}_{j})},    
\end{equation}
where, $\vthetha$ represents the approximate MLP and comprises all the aforementioned parameters, i.e.,  $m^{(l)}_{i,j},s^{(l)}_{i,j},k^{(l)}_{i,j},b^{(l)}_{j}$, $\forall i, j, l$,
and $\vthetha^{(l)}_{j}$ represents the approximate neuron $j$ of layer $l$ and includes all the relevant parameters for that specific neuron.
For each approximate MLP, $\vthetha = \bigcup_{\forall l,j} \vthetha^{(l)}_{j}$.

A simple but effective way to estimate the area of a multi-operand adder is counting the number of Full-Adders (FAs) it instantiates~\cite{weste:2015cmos}. 
For simplicity, we assume only FAs for the reduction.
Each FA performs a 3-to-2 reduction, meaning that for every three bits in a column, one bit remains, and one goes to the column to the right.
Reduction is repeated until only two bits remain in each column.
We implement a Python function to estimate $\operatorname{AdderArea}(\mathbf{\vthetha}^{(l)}_{j})$ that takes as input the weights, masks, and bias of an approximate neuron  $\vthetha^{(l)}_{j}$, i.e., the parameters ${m}^{(l)}_{i,j},{s}^{(l)}_{i,j},{k}^{(l)}_{i,j},{b}^{(l)}_{j}$, $\forall i$, calculates the non-zero bits in each column, and then recursively computes the number of required FAs.


\begin{figure}[!t]
\centering
\includegraphics[scale=0.9]{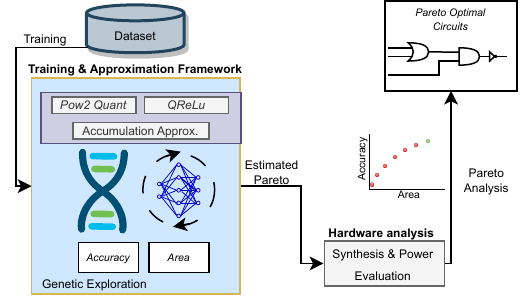}
\vspace{-2ex}
\caption{Abstract high-level overview of our proposed framework. }
\vspace{-3ex}

\label{fig:framework}
\end{figure}

\section{Proposed Framework}\label{sec:framework}

This section describes our framework, an abstract overview of which is depicted in \figurename~\ref{fig:framework}. 
Our framework employs a discrete genetic-based training on a specified MLP topology and dataset to derive an estimated Pareto front of area-accuracy circuits, while applying the approximations discussed in Section~\ref{subsec:ax_methods}.
Subsequently, a hardware evaluation of the evolved MLP circuits is conducted using EDA tools to identify the true Pareto-optimal circuits in terms of area and accuracy.

\subsection{Hardware-Aware Training Flow}\label{subsec:ha-training-flow}

As depicted in \figurename~\ref{fig:framework}, our framework implements a hardware approximation-aware training 
targeting efficient, in terms of accuracy and area, MLP classifiers. 
The training process essentially becomes a multi-objective optimization problem, taking into account both area and accuracy as objectives.
As result, our optimization problem is a multi-objective minimization problem between area and classification error rate defined as:
\begin{equation}
\min_{\vthetha} \; \left[ \operatorname{1-Accuracy}(\vthetha, \mathcal{D}), \, \operatorname{Area}(\vthetha) \right],
\label{eq:problem}
\end{equation}
where, $\mathcal{D}$ denotes the training data and, as defined above, $\vthetha$
includes all learnable parameters $m^{(l)}_{i,j},s^{(l)}_{i,j},k^{(l)}_{i,j}, b^{(l)}_{j}$, $\forall i, j, l$.
These parameters are represented in the discrete domain, which makes it infeasible, especially for the masks, to compute gradients and use traditional backpropagation training.
Therefore, to handle the discrete space we implement a genetic-based training.
The GA will not only explore weights ($s^{(l)}_{i,j},k^{(l)}_{i,j}$) and biases ($b^{(l)}_{j}$) values but will also search the respective masks ($m^{(l)}_{i,j}$) to satisfy \eqref{eq:problem}.
Due to simplicity, low computational complexity, and enhanced convergence, we employ the Non-dominated Sorting Genetic Algorithm II (NSGA-II~\cite{Deb:NSGA:2002}) to address this multi-objective problem.




In the context of employing GA-driven training for MLPs, the weights' update process is contingent upon the genetic operations (i.e., mutation and crossover).
The progress of training is influenced by the evolution of the optimization problem through natural selection, which involves a fitness function.
In MLP training, the mutation operator introduces random alterations to neuron weights, while crossover combines winning weights. These genetic operations are applied randomly during the training process.

The area-accuracy Pareto-optimal points that the GA searches for, are consisted of chromosomes with the best combination of MLP masks, weights, and biases (learnable parameters in $\vthetha$).
When the genetic exploration ends, an estimated area-accuracy Pareto-optimal set is obtained. 
The trained coefficients and masks of the estimated Pareto front, are then automatically translated into an HDL description as discussed above and illustrated for example in \figurename~\ref{fig:ax:neuron}.
Next, hardware analysis is employed and the true Pareto front among the evaluated designs is obtained.

In order to facilitate the convergence of the evolutionary algorithm to a Pareto-optimal set, we create an initial population of semi-random chromosomes.
This population is randomly selected and further doped with a small percentage (\textapprox $10$\%) of nearly non-approximate solutions, exploring solutions of high accuracy at the early stages of evolution.
Additionally, we impose a $10$\% upper bound on the acceptable accuracy loss, compared to the baseline accuracy~\cite{Mubarik:MICRO:2020:printedml}, to consider solutions during the training that exhibit small accuracy degradation.

\begin{figure}[!t]
\centering
\includegraphics[width=\columnwidth]{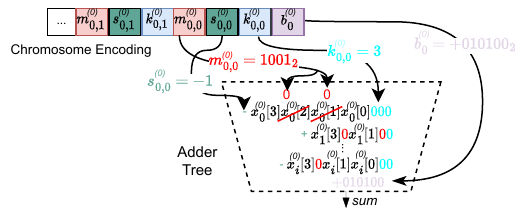}
\vspace{-4.2ex}
\caption{\red{Chromosome encoding and neuron construction.}}
\label{fig:chromosome} 
\vspace{-3.5ex}
\end{figure}

\subsection{Formulation and Encoding of the Approximate Neuron}\label{subsec:ha-formulation}

The approximations embedded during the training include pow2 weight quantization and fine-grain unstructured pruning.
Integrating pow2 quantization into training simply restricts the solution space of the MLP's coefficients within pow2 values.
As for the pruning method, we introduce an additional training parameter, i.e., the masks.

Treating the masks as a training parameter means that the corresponding decisions are incorporated into the exploration during training.
Therefore, the output of the $j$-th neuron in the $l$-th layer of the hardware-approximated MLP is given by:

\begin{equation}\label{eq:masks}
\operatorname{QReLU}\left(\sum_{i}{s^{(l)}_{i,j} ({m^{(l)}_{i,j}\odot x^{(l)}_i)\!\ll\! k^{(l)}_{i,j}}}+b^{(l)}_j\right),
\end{equation}
where, $\ll$ is the bit-shift operator, and $x^{(l)}_i$, $\forall i$, are input activations of the $l$-th layer.
The rest are our learnable parameters.



Our genetic-based training must identify all the learnable parameters in $\vthetha$ that minimize our objective function~\eqref{eq:problem}.
The evaluated fitness function involves calculating the inference accuracy using~\eqref{eq:masks} and estimating the hardware overhead using~\eqref{eq:area} and our high-level FA-count Python function. 

Each gene in the GA chromosome represents a $m^{(l)}_{i,j}$, $s^{(l)}_{i,j}$, $k^{(l)}_{i,j}$, or $b^{(l)}_{j}$ parameter. 
An example of a chromosome's encoding is illustrated in \figurename~\ref{fig:chromosome} where the genes are grouped by weight ($m^{(l)}_{i,j}$, $s^{(l)}_{i,j}$, $k^{(l)}_{i,j}$) then by neuron, and finally by layer.
Hence each gene is represented by an integer value (with the corresponding limits, e.g., weight size for $k^{(l)}_{i,j}$).
Furthermore, \figurename~\ref{fig:chromosome} shows the direct relationship between the genes, the applied approximations, and the eventual hardware implementation, which highlights the hardware-awareness of our approach.

\section{Results and Evaluation}\label{sec:experimental}

In this section, we conduct a comprehensive evaluation of our framework. 
We evaluate the area and accuracy of the printed MLPs trained with our framework and compare them against the state-of-the-art exact baseline~\cite{Mubarik:MICRO:2020:printedml} as well as the proposed state-of-the-art printed MLPs~\cite{Armeniakos:TCAD2023:cross, Armeniakos:TC2023:codesign, Weller:2021:printed_stoch}.
Finally, we assess the effectiveness of our framework in enabling printed-battery-powered MLP classifiers.

\subsection{Experimental Setup}\label{subsec:exp_setup}

We examine five datasets, namely Breast Cancer (BC), Cardiotocography (Ca), Pendigits (PD), Red Wine (RW), and White Wine (WW), from~\cite{Dua:2019:uci}. 
These datasets could form realistic printed applications as they feature inputs  suitable for printed circuits and demand low precision, duty cycle, and sample rate requirements~\cite{Bleier:ISCA:2020:printedmicro}.
They have also been previously utilized in the related works \cite{Mubarik:MICRO:2020:printedml, Armeniakos:TC2023:codesign, Armeniakos:TCAD2023:cross, Weller:2021:printed_stoch}, ensuring a fair comparison for our analysis.

The inputs are normalized to $[0, 1]$ as in~\cite{Mubarik:MICRO:2020:printedml, Armeniakos:TC2023:codesign} and are randomly stratified split into $70\%/30\%$ train/test sets, ensuring a balanced distribution of each target class in each of these sets.
Mutation and crossover operators of the GA are set to $0.2\%$ \text{and} $0.7\%$ respectively.
All circuits are synthesized using Synopsys Design Compiler S-2021.06 and mapped to the printed EGFET library~\cite{Bleier:ISCA:2020:printedmicro}, while VCS T-2022.06 and PrimeTime T-2022.03 are used for simulation and power analysis.
The accuracy is reported on the test dataset, and all designs are synthesized at a relaxed clock period to improve area efficiency.
Clock period of $200\si{\milli\second}$ are applied to all MLPs, except for Pendigits, which requires $250\si{\milli\second}$.
Such delay values align with typical PE performance~\cite{cadilha2017digital}.
The architecture of the MLPs is the same as the authors have reported in~\cite{Mubarik:MICRO:2020:printedml} and~\cite{Armeniakos:TC2023:codesign}.
Our exact baseline are the bespoke printed MLPs circuits, designed following the approach outlined in~\cite{Mubarik:MICRO:2020:printedml}, using $8$-bit fixed point weights and $4$-bit inputs.
Their hardware characteristics, accuracy, and topology are reported in Table~\ref{tab:baselines}.
Given the large hardware overheads in printed circuits and the feasibility constraints of printed MLPs, we consider a $5\%$ accuracy loss compared to the exact baseline~\cite{Mubarik:MICRO:2020:printedml} (see Table~\ref{tab:baselines}) as an acceptable level of accuracy for our experiments.

\begin{table}[t!]
\setlength\tabcolsep{3pt}
\caption{Evaluation of our Printed MLP for up to 5\% Accuracy Loss.}
\label{tab:res1v}
\footnotesize
\centering
\renewcommand{\arraystretch}{1.1}
\begin{threeparttable}
\begin{tabular}{|l|ccc|cc|}
\cline{2-6}
 \multicolumn{1}{c|}{}  & \multicolumn{5}{c|}{\textbf{Our Approximate MLPs}}  \\ \hline

\multicolumn{1}{|c|}{\textbf{MLP}} &  \textbf{Accuracy}    & \begin{tabular}[c]{@{}c@{}}\textbf{Area} \\ ($\si{\square\centi\meter}$)\end{tabular} & \begin{tabular}[c]{@{}c@{}}\textbf{Power}\\ ($\si{\milli\watt}$)\end{tabular}  & \makecell{\textbf{Area} \\ \textbf{Reduction}\tnote{1}} & \makecell{\textbf{Power} \\ \textbf{Reduction}\tnote{1}} \\ \hline
  
\textbf{Breast Cancer}     & 0.947 & 0.04  & 0.15  &288\(\times\)    &274\(\times\)    \\
\textbf{Cardio}           & 0.873 & 1.73  & 6.5   &19.3\(\times\)   &19.0\(\times\)  \\
\textbf{Pendigits}        & 0.893 & 12.7  & 40.2  &5.3\(\times\)    &5.3\(\times\)    \\
\textbf{RedWine}             & 0.519 & 0.04  & 0.13  &470\(\times\)    &579\(\times\)    \\
\textbf{WhiteWine}           & 0.508 & 0.20  & 0.74  &122\(\times\)    &137\(\times\)    \\ \hline
\end{tabular}
\begin{tablenotes}\footnotesize
\item[] 
$^1$ With respect to the corresponding bespoke exact baseline\cite{Mubarik:MICRO:2020:printedml}.
\vspace{-3ex}
\end{tablenotes}
\end{threeparttable}
\vspace{-1.3ex}
\end{table}



\subsection{Comparison Against the Baseline and State of the Art}\label{subsec:comparisons}

First, we evaluate our framework and the state-of-the-art exact baseline MLPs~\cite{Mubarik:MICRO:2020:printedml}.
Table~\ref{tab:res1v} shows the area, power, and accuracy of our printed MLP that feature up to $5\%$ accuracy loss.
As shown, compared to the baseline, our MLP circuits achieve \red{$181\times$} and \red{$203\times$} area and power reduction on average, respectively.
Notably the area gains range from \red{$5.3\times$} to \red{$470\times$} while the power gains range from \red{$5.3\times$} to \red{$578\times$}.

\begin{figure}[!t]
\centering
\includegraphics[width=\columnwidth]{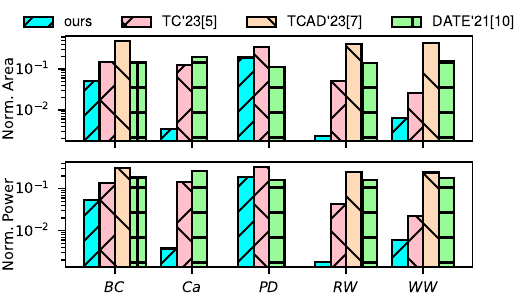}
\vspace{-4ex}
\caption{\red{(a) Area and (b) power reduction delivered by our printed MLPs and the state-of-the-art approximate~\cite{Armeniakos:TCAD2023:cross,Armeniakos:TC2023:codesign} and stochastic~\cite{Weller:2021:printed_stoch} ones.
Area and power are normalized w.r.t. the baseline exact MLPs~\cite{Mubarik:MICRO:2020:printedml}.
 Y-axis is in logarithmic scale.}}
\label{fig:compsoa}
\vspace{-3.5ex}
\end{figure}

In \figurename~\ref{fig:compsoa}, we present a comparison of the area and power gains of our printed MLPs compared with the state-of-the-art approximate~\cite{Armeniakos:TCAD2023:cross, Armeniakos:TC2023:codesign} and stochastic~\cite{Weller:2021:printed_stoch} ones.
All MLPs in \figurename~\ref{fig:compsoa} feature the same inference latency.
Despite a small clock period in~\cite{Weller:2021:printed_stoch}, each inference takes $220$-$230\si{\milli\second}$ due to the use of a stochastic bitstream of length $1024$.
For our circuits and~\cite{Armeniakos:TC2023:codesign, Weller:2021:printed_stoch} a $1\si{\volt}$ operation is considered.
In~\cite{Armeniakos:TCAD2023:cross}, Voltage Over-Scaling is used and the MLPs are operated below $0.8\si{\volt}$.
In \figurename~\ref{fig:compsoa}, all values are normalized over the corresponding exact bespoke design~\cite{Mubarik:MICRO:2020:printedml}.
Although, we and the authors in~\cite{Armeniakos:TC2023:codesign,Armeniakos:TCAD2023:cross} consider up to $5$\% accuracy loss compared to the baseline~\cite{Mubarik:MICRO:2020:printedml}, it's worth noting that \cite{Weller:2021:printed_stoch} cannot achieve such high accuracy.
In fact, the average accuracy loss for the MLPs they consider in~\cite{Weller:2021:printed_stoch} is $35$\%.

As shown in Fig.~\ref{fig:compsoa}, our framework significantly outperforms~\cite{Armeniakos:TC2023:codesign},~\cite{Armeniakos:TCAD2023:cross} and~\cite{Weller:2021:printed_stoch}.
Specifically, compared to~\cite{Armeniakos:TC2023:codesign}, our MLPs achieve on average \red{$13\times$} area reduction and \red{$14\times$} power reduction.
The area and power gains range from \red{$1.8\times$} to \red{$36\times$} and from \red{$1.8\times$} to \red{$39\times$}, respectively.
Similarly, compared to~\cite{Armeniakos:TCAD2023:cross}, our MLPs achieve \red{$25\times$} lower area and \red{$14.5\times$} lower power on average.
In~\cite{Armeniakos:TCAD2023:cross} Pendigits is not considered, possible due to its high complexity.
Finally, our MLPs deliver \red{$19\times$} and \red{$26\times$} area and power saving, respectively, compared to~\cite{Weller:2021:printed_stoch}.
Only for Pendigits, the stochastic MLP of~\cite{Weller:2021:printed_stoch} attains slightly lower power and area than our approximate MLP.
Though,~\cite{Weller:2021:printed_stoch} achieves only $22\%$ accuracy while we achieve \red{$89.3\%$}.

\subsection{Printed-Battery Power Operation}\label{subsec:exp3}

Next, we evaluate the effectiveness of our framework, in generating printed-battery powered MLP classifiers.
As shown in Table~\ref{tab:res1v}, Breast Cancer, RedWine, and WhiteWine classifiers are compatible with a Blue Spark $5 \si{\milli\watt}$ battery, while Cardio can be powered by a Zinergy $15 \si{\milli\watt}$ battery.
Pendigits, with its larger topology requiring $145$ parameters, cannot be powered even with a Molex $30 \si{\milli\watt}$ battery.

Our applied approximations result in faster MLPs compared to their exact baseline equivalents.
This enables further power reduction by scaling the supply voltage without sacrificing performance, i.e., maintaining the same latency as the baseline circuit~\cite{Mubarik:MICRO:2020:printedml}.
Hence, we can now power previously unpowered MLPs like Pendigits or use smaller printed energy sources.
This is very advantageous for compact, self-powered designs like wearables. 
Implantable and wearable medical devices prioritize functionality, user comfort, and device longevity, thus harvesting energy from the body is highly efficient. Integrating energy harvesting ensures continuous operation~\cite{shuvo:energy2022:harvesting}.

\begin{figure}[!t]
\centering
\includegraphics[scale=0.9]{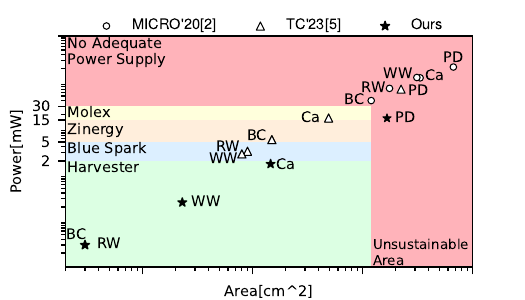}
\vspace{-2.3ex}
\caption{Feasibility evaluation.
Categorizing our printed MLPs, the baseline ~\cite{Mubarik:MICRO:2020:printedml} and the approximate~\cite{Armeniakos:TC2023:codesign} ones based on their sustainability concerning area overhead and the availability of printed power sources.}
\label{fig:bateries}
\vspace{-3.2ex}
\end{figure}

Considering that EGFET printed circuits can operate down to $0.6\si{\volt}$~\cite{Marques:Materials:2019} and that printed batteries are customizable in terms of polarity, voltage, shape, etc.~\cite{PrintedBatteries2018}, we set the voltage supply of our approximate MLPs (MLPs in Table~\ref{tab:res1v}) to the minimum supported value, i.e., $0.6\si{\volt}$, and re-synthesize our designs.
In \figurename~\ref{fig:bateries}, our MLPs, along with the baseline \cite{Mubarik:MICRO:2020:printedml} and~\cite{Armeniakos:TC2023:codesign}, are categorized based on their area and suitable power source.
Again, the $5\%$ accuracy constraint is considered.
As shown in \figurename~\ref{fig:compsoa}, among the state-of-the-art works,~\cite{Armeniakos:TC2023:codesign} outperformed~\cite{Armeniakos:TCAD2023:cross,Weller:2021:printed_stoch} in area-power gains and accuracy.
In \figurename~\ref{fig:bateries}, the red zone denotes an unsustainable area where the circuit's size is considered excessively large for most printed applications or where there isn't an adequate power supply.
The other zones use a different color based on the printed power source considered, e.g., printed battery Molex $30\si{\milli\watt}$.

\figurename~\ref{fig:bateries} demonstrates how our framework benefits the area-power trade-off by shifting the MLPs to a lower area-power space. 
As shown, all the baseline MLPs lie in the red zone while the approximate MLPs~\cite{Armeniakos:TC2023:codesign} mainly require large batteries.
On the other hand, all our circuits, except for Pendigits lie within the green zone, indicating that they can be powered by only a printed energy harvester.
Although our Pendigits MLP can be now powered by a Molex $30\si{\milli\watt}$ battery, its area might be impractical for most printed applications.
Note that the Pendigits MLPs of~\cite{Armeniakos:TC2023:codesign} and~\cite{Mubarik:MICRO:2020:printedml} cannot be powered by any existing printed power source.
On average, our MLPs at $0.6\si{\volt}$ achieve $\red{912\times}$ lower power compared to the baseline~\cite{Mubarik:MICRO:2020:printedml}. 
Similarly, compared to~\cite{Armeniakos:TC2023:codesign}, our MLPs at $0.6\si{\volt}$ achieve $\red{65\times}$ lower power on average.

\subsection{Execution Time Evaluation}


Table~\ref{tab:timings} presents training execution time on an AMD EPYC 7552 with $256$GB RAM. 
As expected, gradient-based training is faster than \blue{the conventional} for the same number of evaluations, i.e., without approximation and hardware-awareness, GA-based training. 
Our approximate hardware-aware GA-based training averages only $100$ minutes, even with over $26$ million chromosome evaluations. 
This time is close to hardware-unaware conventional GA-based training.
\blue{
The runtime impact is minimal, albeit integrating the addition approximation into the training process, doubles the trainable parameters, necessitating a mask to be trained for each neuron's input.
}
\blue{
Overall, our approach features reasonable execution time, especially when considering the addressed limitations and complexity 
}
\vspace{-3px}
\section{Conclusion}

Printed electronics offer cost-effective, flexible, and conformal hardware, making it ideal for ultra-low-cost applications. 
However, the associated large feature size limits the feasibility of complex ML classifiers like MLPs.
To overcome this, we integrate hardware approximation into the training and introduce a GA-based hardware-aware training method to design approximate bespoke printed MLP circuits.
Our approach efficiently explores the discrete hardware approximation space, striking a balance between accuracy and hardware efficiency.
Our evaluation shows that our printed MLPs outperform the \blue{state of the art} in terms of hardware efficiency while maintaining similar accuracy.
This advancement enables printed-battery-powered operation for all examined MLPs, with most of them being self-powered using a printed energy harvester.



{\small
\section*{Acknowledgment}
This work is partially supported by the European Research Council (ERC) and co-funded by the H.F.R.I call “Basic research Financing (Horizontal support of all Sciences)” under the National Recovery and Resilience Plan “Greece 2.0” (H.F.R.I. Project Number: 17048).
}
\begin{table}[t!]
\setlength\tabcolsep{3pt}
\caption{Training Execution Times Evaluation in Minutes}
\label{tab:timings}
\footnotesize
\centering
\renewcommand{\arraystretch}{1}
\begin{threeparttable}
\begin{tabular}{|l|cc|c|}
\cline{1-4}


\multicolumn{1}{|c|}{\textbf{MLP}} & \makecell{\textbf{Exec.Time} \\ \textbf{Grad.} (min)\textbf{\tnote{1}}}  & \makecell{\textbf{Exec.Time} \\ \textbf{GA} (min)\textbf{\tnote{2}}}  & \makecell{\textbf{Exec.Time} \\ \textbf{GA-AxC} (min)\textbf{\tnote{3}}}\\ \hline
  
\textbf{Breast Cancer}      & 0.5&  8  & 9 \\
\textbf{Cardio}             & 2  &  42 & 45\\
\textbf{Pendigits}          & 14 &  298 & 344 \\
\textbf{RedWine}            & 2  &  21  & 22\\
\textbf{WhiteWine}          & 7  &  77  & 79\\
\cline{1-4}
\textbf{Average}            & 5  &  89  & 100\\ \hline

\end{tabular}
\begin{tablenotes}\footnotesize
\item[] 
$^1$ Gradient with only accuracy as objective.
$^2$ GA-based with only accuracy as objective.
$^3$ GA-based with AxC techniques and both accuracy and area as objectives.

\vspace{-4ex}
\end{tablenotes}
\end{threeparttable}
\end{table}


\vspace{-3ex}

\bibliographystyle{IEEEtran}
\bibliography{IEEEabrv,references}

\begin{thebibliography}{10}
\providecommand{\url}[1]{#1}
\csname url@samestyle\endcsname
\providecommand{\newblock}{\relax}
\providecommand{\bibinfo}[2]{#2}
\providecommand{\BIBentrySTDinterwordspacing}{\spaceskip=0pt\relax}
\providecommand{\BIBentryALTinterwordstretchfactor}{4}
\providecommand{\BIBentryALTinterwordspacing}{\spaceskip=\fontdimen2\font plus
\BIBentryALTinterwordstretchfactor\fontdimen3\font minus \fontdimen4\font\relax}
\providecommand{\BIBforeignlanguage}[2]{{%
\expandafter\ifx\csname l@#1\endcsname\relax
\typeout{** WARNING: IEEEtran.bst: No hyphenation pattern has been}%
\typeout{** loaded for the language `#1'. Using the pattern for}%
\typeout{** the default language instead.}%
\else
\language=\csname l@#1\endcsname
\fi
#2}}
\providecommand{\BIBdecl}{\relax}
\BIBdecl

\bibitem{Bleier:ISCA:2020:printedmicro}
N.~Bleier, M.~Mubarik, F.~Rasheed, J.~Aghassi-Hagmann, M.~B. Tahoori, and R.~Kumar, ``Printed microprocessors,'' in \emph{Annu. Int. Symp. Computer Architecture (ISCA)}, jun 2020, pp. 213--226.

\bibitem{Mubarik:MICRO:2020:printedml}
M.~H. Mubarik \emph{et~al.}, ``Printed machine learning classifiers,'' in \emph{Annu. Int. Symp. Microarchitecture (MICRO)}, 2020, pp. 73--87.

\bibitem{cadilha2017digital}
G.~Cadilha~Marques \emph{et~al.}, ``Digital power and performance analysis of inkjet printed ring oscillators based on electrolyte-gated oxide electronics,'' \emph{Applied Physics Letters}, vol. 111, no.~10, p. 102103, 2017.

\bibitem{lei2019low}
T.~Lei \emph{et~al.}, ``Low-voltage high-performance flexible digital and analog circuits based on ultrahigh-purity semiconducting carbon nanotubes,'' \emph{Nature communications}, vol.~10, no.~1, p. 2161, 2019.

\bibitem{Armeniakos:TC2023:codesign}
G.~Armeniakos, G.~Zervakis, D.~Soudris, M.~B. Tahoori, and J.~Henkel, ``Co-design of approximate multilayer perceptron for ultra-resource constrained printed circuits,'' \emph{{IEEE} Trans. Comput.}, pp. 1--8, 2023.

\bibitem{Armeniakos:DATE2022:axml}
G.~Armeniakos, G.~Zervakis, D.~Soudris, M.~B. Tahoori, and J.~Henkel, ``Cross-layer approximation for printed machine learning circuits,'' in \emph{Design Automation and Test in Europe (DATE)}, 2022, pp. 190--195.

\bibitem{Armeniakos:TCAD2023:cross}
G.~Armeniakos, G.~Zervakis, D.~Soudris, M.~B. Tahoori, and J.~Henkel, ``Model-to-circuit cross-approximation for printed machine learning classifiers,'' \emph{{IEEE} Trans. Comput.-Aided Design Integr. Circuits Syst.}, pp. 1--1, 2023.

\bibitem{Kokkinis:DATE2023}
A.~Kokkinis \emph{et~al.}, ``Hardware-aware automated neural minimization for printed multilayer perceptrons,'' in \emph{Design Automation and Test in Europe (DATE)}, 2023.

\bibitem{Henkel:ICCAD2022:expedition}
J.~Henkel \emph{et~al.}, ``Approximate computing and the efficient machine learning expedition,'' in \emph{Int. Conf. on Computer-Aided Design (ICCAD)}, 2022, pp. 1--9.

\bibitem{Weller:2021:printed_stoch}
D.~D. Weller \emph{et~al.}, ``Printed stochastic computing neural networks,'' in \emph{Design Automation and Test in Europe (DATE)}, 2021, pp. 914--919.

\bibitem{Armeniakos:AxDNNsurvey}
G.~Armeniakos, G.~Zervakis, D.~Soudris, and J.~Henkel, ``Hardware approximate techniques for deep neural network accelerators: A survey,'' \emph{ACM Comput. Surv.}, vol.~55, no.~4, nov 2022.

\bibitem{benmeziane:arxiv2021:hw_awareNAS}
\BIBentryALTinterwordspacing
H.~Benmeziane, K.~El~Maghraoui, H.~Ouarnoughi, S.~Niar, M.~Wistuba, and N.~Wang, \emph{{A Comprehensive Survey on Hardware-Aware Neural Architecture Search}}, Jan. 2021. [Online]. Available: \url{https://uphf.hal.science/hal-03269441}
\BIBentrySTDinterwordspacing

\bibitem{sekanina:TEvolutioryC:2015:evolution_to_ax_circuits}
Z.~Vasicek and L.~Sekanina, ``Evolutionary approach to approximate digital circuits design,'' \emph{{IEEE} Trans. Evol. Comput.}, vol.~19, no.~3, pp. 432--444, 2015.

\bibitem{cui2016printed}
Z.~Cui, \emph{Printed electronics: materials, technologies and applications}.\hskip 1em plus 0.5em minus 0.4em\relax John Wiley \& Sons, 2016.

\bibitem{chang2017circuits}
J.~S. Chang, A.~F. Facchetti, and R.~Reuss, ``A circuits and systems perspective of organic/printed electronics: review, challenges, and contemporary and emerging design approaches,'' \emph{{IEEE} J. Emerg. Sel. Top. Circuits Syst.}, vol.~7, no.~1, pp. 7--26, 2017.

\bibitem{weste:2015cmos}
N.~H. Weste and D.~Harris, \emph{CMOS VLSI design: a circuits and systems perspective}.\hskip 1em plus 0.5em minus 0.4em\relax Pearson Education India, 2015.

\bibitem{Deb:NSGA:2002}
D.~Kalyanmoy, P.~Amrit, A.~Sameer, and T.~Meyarivan, ``A fast and elitist multiobjective genetic algorithm: Nsga-ii,'' \emph{{IEEE} Trans. Evol. Comput.}, vol.~6, no.~2, pp. 182--197, 2002.

\bibitem{Dua:2019:uci}
D.~Dua and C.~Graff, ``{UCI} machine learning repository,'' 2017.

\bibitem{shuvo:energy2022:harvesting}
M.~M.~H. Shuvo, T.~Titirsha, N.~Amin, and S.~K. Islam, ``Energy harvesting in implantable and wearable medical devices for enduring precision healthcare,'' \emph{Energies}, vol.~15, no.~20, p. 7495, 2022.

\bibitem{Marques:Materials:2019}
C.~Marques \emph{et~al.}, ``{Progress Report on “From Printed Electrolyte-Gated Metal-Oxide Devices to Circuits”},'' \emph{Advanced Materials}, vol.~31, 2019.

\bibitem{PrintedBatteries2018}
S.~Lanceros‐Méndez and C.~M. Costa, \emph{Printed Batteries: Materials, Technologies and Applications}.\hskip 1em plus 0.5em minus 0.4em\relax Wiley, 2018.

\end{thebibliography}

\end{document}